# Breaking Imphash

Chris Balles
SCYTHE, Inc.
chris@scythe.io

Ateeq Sharfuddin
SCYTHE, Inc.
ateeq@scythe.io

***Abstract*—There are numerous schemes to generically signature artifacts. We specifically consider how to circumvent signatures based on imphash. Imphash is used to signature Portable Executable (PE) files and an imphash of a PE file is an MD5 digest over all the symbols that PE file imports. Imphash has been used in numerous cases to accurately tie a PE file seen in one environment to PE files in other environments, although each of these PE files' contents was different. An argument made for imphash is that alteration of imphashes of derived PE file artifacts is unlikely since it is an expensive process, such that you will need to either modify the source code and recompile or relink in a different order. Nevertheless, we present a novel algorithm that generates derivative PE files such that its imphash is different from the original PE file. This straightforward algorithm produces feasible solutions that defeat approaches relying on the impash algorithm to signature PE files.**

## I. Introduction

Incident Response and infrastructure teams want to efficiently detect, disrupt, and prevent threat groups running malware on enterprise devices. Tracking a threat group's file artifacts assists in identifying their attack campaigns. One well-known scheme used to track threat groups' backdoors is imphash [1]. Imphash (for "import hash") is a signature scheme that identifies portable executable (PE) files' imports uniquely, and has been used in numerous cases (e.g., [2][3][6][7][8]) to accurately tie a threat group's backdoor PE file seen in one environment to backdoor PE files in other environments even though each of these files' contents was different. Although the commonly-held belief is that any solution to avoid imphash signatures is time-consuming and expensive for malware authors, necessitating malware authors to modify source code and recompile or relink in a different order to generate variants, we present a novel algorithm that generates derivative PE files each having an imphash different from the original PE file. This proposed algorithm demonstrates that using imphash schemes to track threat groups' PE files is not viable.

### A. Related Work

Information Technology (IT) and Incident Response (IR) teams who defend enterprise systems prefer signature schemes that are inexpensive to compute but prohibitively expensive for malware authors to defeat. Similarly, malware authors prefer signature avoidance techniques that inexpensively transform the original file artifact but are prohibitively expensive for defenders to signature.

There are multiple schemes to signature artifacts. A straightforward approach is to produce a message digest, such as MD5, SHA1, SHA256, etc., over the content of a file. Any change to the file's content yields a new message digest, and this new message digest has no similarity to the original file's message digest. There are locality-sensitive hashing algorithms that produce hashes such that the distance between the hashes will be small if the files' contents are similar to each other. Other approaches to signature PE files involve checking if the program database (pdb) path are the same or if the timestamp the linker stored in the file is the same. Imphash, as mentioned previously, produces an MD5 digest over all the symbols a PE file imports. There are numerous papers reporting usage of imphash to cluster malware to same families [2][3].

### B. Portable Executable

Microsoft designed the Portable Executable (PE) file format for use in all their operating systems [4]. This format is loosely documented in WINNT.H and thoroughly documented in [5], and describes all the pieces necessary to run code encapsulated inside a file employing this format. Windows executables such as EXEs, DLLs, SYSs, etc. all are files following this format. The Import Table, the Import Address Table, and the Base Relocation Table are of particular interest.

Each PE file begins with an IMAGE_DOS_HEADER structure, which contains the offset to the IMAGE_NT_HEADERS structure. The IMAGE_OPTIONAL_HEADER structure in IMAGE_NT_HEADERS contains data directories storing offsets to the import table, import address table, and the base relocation table in the PE file.

#### 1) Import Table

The import table describes all the symbols the PE file uses implicitly and the modules in which these symbols reside. For example, if your PE file great.exe calls the function ole32!CoInitialize implicitly, the import table must contain a reference to the module ole32.dll and the symbol CoInitialize. When the image of great.exe is loaded for execution, the loader shall implicitly load ole32.dll to ensure that the call to ole32!CoInitialize by great.exe succeeds.

The import table is an array of image descriptors, one for each module the PE implicitly links. The image import descriptor contains an offset to the name of the module (e.g., ole32.dll) to be imported, as well as the offset to an original first thunk and the first thunk. An IMAGE_THUNK_DATA structure represents each thunk, and if IMAGE_ORDINAL_FLAG is not set in the thunk, IMAGE_THUNK_DATA is an offset to an IMAGE_IMPORT_BY_NAME structure containing the symbol name (e.g., CoInitialize) to be imported.

```
struct IMAGE_IMPORT_DESCRIPTOR
{
   union
   {
      DWORD   Characteristics;
      DWORD   OriginalFirstThunk;
   };

   DWORD   TimeDateStamp
   DWORD   ForwarderChain;
   DWORD   Name;
   DWORD   FirstThunk;
};

union IMAGE_THUNK_DATA
{
   DWORD ForwarderString;
   DWORD Function;
   DWORD Ordinal;
   DWORD AddressOfData;
};

struct IMAGE_IMPORT_BY_NAME
{
   WORD Hint;
   CHAR Name[1];
};
```

Code 1: Import table structures

*2) Import Address Table*

The content of the import address table is identical to the import table. However, during binding, the loader updates the entries in the import address table to the actual virtual memory addresses of the imported symbols. The FirstThunk field in IMAGE_IMPORT_DESCRIPTOR points to the first thunk in this import address table.

*3) Base Relocation Table*

The base relocation table contains blocks of entries that need to be updated should the PE not be loaded at its desired base address.

## II. Algorithms and Implementation

### *A. Imphash algorithm*

The earliest references to Imphash appear to be in [1] and [6]. Imphash is now widely applied and used to cluster similar malware [7]. To generate imphash, iterate over the import table and append all the symbols for each module to be imported as module.symbol (lowercase) into a string ordered as iterated. Finally, compute the MD5 of this string. A simple best-case imphash function is presented in pseudocode below.

```
function get_imphash():
        str = “”
        for entry in DIRECTORY_ENTRY_IMPORT:
          module_name = lower_case(entry.name).strip_extension()
          for import in entry.imports:
            function = import.name
            str += module_name + “.” + lower_case(function)
        return md5(str)
```

Code 2: Imphash algorithm pseudocode (without exceptional cases)

This imphash algorithm allows defenders to easily tie all the attack campaigns’ PE file artifacts derived from one original PE file artifact. Simply adding or updating resources or any other data in the PE which do not affect implicitly loaded symbols does not change the order of the imports. Therefore, this computed imphash is the same for all of these PE artifacts. The only way for imphash to be different is if the order in which the imported functions were used in code had changed, resulting in the linker generating a different reordering of the imported symbols. Consequently, this makes it expensive for malware authors: to generate an artifact to have a different imphash, have to reorder the function calls, maybe recompile, and relink as described in [1]. The argument made in support of imphash [1] is that this algorithm is expensive to defeat and alludes that there are not many malware authors to care enough to write tools to modify imphash. This second conclusion is weak, and the algorithm presented next trivially demonstrates how numerous file artifacts can be derived, each with an imphash different from the original PE file.

### *B. Breaking-Impash algorithm*

Given a PE file $E$ with imphash $e$, transform $E$ to a new PE file $E_T$ such that $E_T$ now has imphash $e_T$, where $e \neq e_T$ but functionally, $E$ and $E_T$ must be one and the same. The algorithm shall transform the PE file without any requirement for access to the source code, or program database, or require recompilation or relinking.

```
thunk_data = []       # vector of (orig_thunk.addressofdata, thunk)
thunk_data_copy = []  # holds duplicate of thunk_data
original_to_new = {}  # map<firstthunk, firstthunk>

for entry in DIRECTORY_ENTRY_IMPORT:
  for import in entry.imports:
     thunk_data.append((import.original_thunk.addressofdata,
                 import.firstthunk))

thunk_data_copy = thunk_data.clone()
random_shuffle(thunk_data)
for i in range(thunk_data.length):
  original_to_new[thunk_data[i].second] = thunk_data_copy[i].second

for entry in DIRECTORY_ENTRY_IMPORT:
  i = 0
  for import in entry.imports:
     import.original_thunk.addressofdata =
        import.firstthunk.addressofdata = thunk_data[i].first
     i = i+1

for relocation_block in DIRECTORY_ENTRY_BASERELOC:
   for patch_address in relocation_block:
      lookup = patch_address - imagebase
      value = original_to_new.find(lookup)
      if value:
         relocation_block[patch_address] = value + imagebase
```

Code 3: Breaking-Imphash algorithm pseudocode

The Breaking-Impash algorithm iterates over the import table and for each module stores a pair (OriginalThunk.AddressOfCode, Thunk) into an array $A$ for each symbol’s original thunk and thunk. The algorithm duplicates this array $A$ to $A_T$ and randomizes the order of the elements in $A_T$. The algorithm then iterates over the import table again, and this time updates the original thunks and thunks to the values in $A_T$. Updating the thunks updates the import address table. Updating the import table and import address table is not sufficient: Code references the thunks and expects the corresponding virtual memory addresses for the imported symbols to be in those thunks. Therefore, after the randomization step for $A_T$. the algorithm stores a mapping of $A_T$ to $A$ in a map $M$. When the entire import table and import address table have been processed the algorithm iterates over the entries in the base relocation table, searching for these addresses as keys in the map $M$. If an address is found in $M$, the algorithm replaces the value with corresponding map value. The saved derived PE file artifact now behaves exactly as the original PE file artifact but has a different imphash.

## III. Result

### A. Time-complexity

Assuming there are $n$ symbols, the cost to randomize and update the thunks is linear and thus, $O(n)$. We also need to store a mapping $M$ between the thunk original offsets and new offsets. Let's assume $M$ is a red-black tree. Therefore, storing $n$ offsets shall be $O(n \lg n)$. We will update the base relocation table, which contains m entries. Therefore, searching $m$ entries in $M$ shall cost $O(m \lg n)$. Our total cost is: $O(n \lg n) + O(m \lg n) + O(n)$. The expectation is that $m >> n$, as more addresses exist needing relocation than just the $n$ thunks. Therefore, this algorithm is bounded $O(m \lg n)$.

Assuming there are $n$ symbols, we can have $n!$ possible orderings. So, if we randomly choose an order following Gaussian distribution, the chances of a PE having same imphash as previous is $1/n!$.

### B. Demonstration with example

For demonstration, we construct an example PE file, great.exe, which implicitly imports three functions CoInitialize, CoCreateGuid, and CoUninitialize from ole32.dll, and kernel32.dll. In a successful execution, the implementation calls CoInitialize, CoCreateGuid, and CoUninitialize in sequence.

```
#include <stdio.h>
#include <windows.h>

int main() {
   HRESULT hr = CoInitialize(NULL);
   GUID = { 0 };
   CoCreateGuid(&guid);
   printf("Hello, World! %x\n", guid.Data1);
   if (SUCCEEDED(hr))
      CoUninitialize();
   return 0;
}
```

Code 4: great.exe source code

The imphash for this great.exe is: 5dc63c6fd3a8ce23292cbe13b6713f16. The IMAGE IMPORT DESCRIPTOR for ole32.dll is shown below in CFF Explorer for great.exe. Notice that the order of the symbols in the descriptor is: CoUninitialize, CoCreateGuid, and CoInitialize.

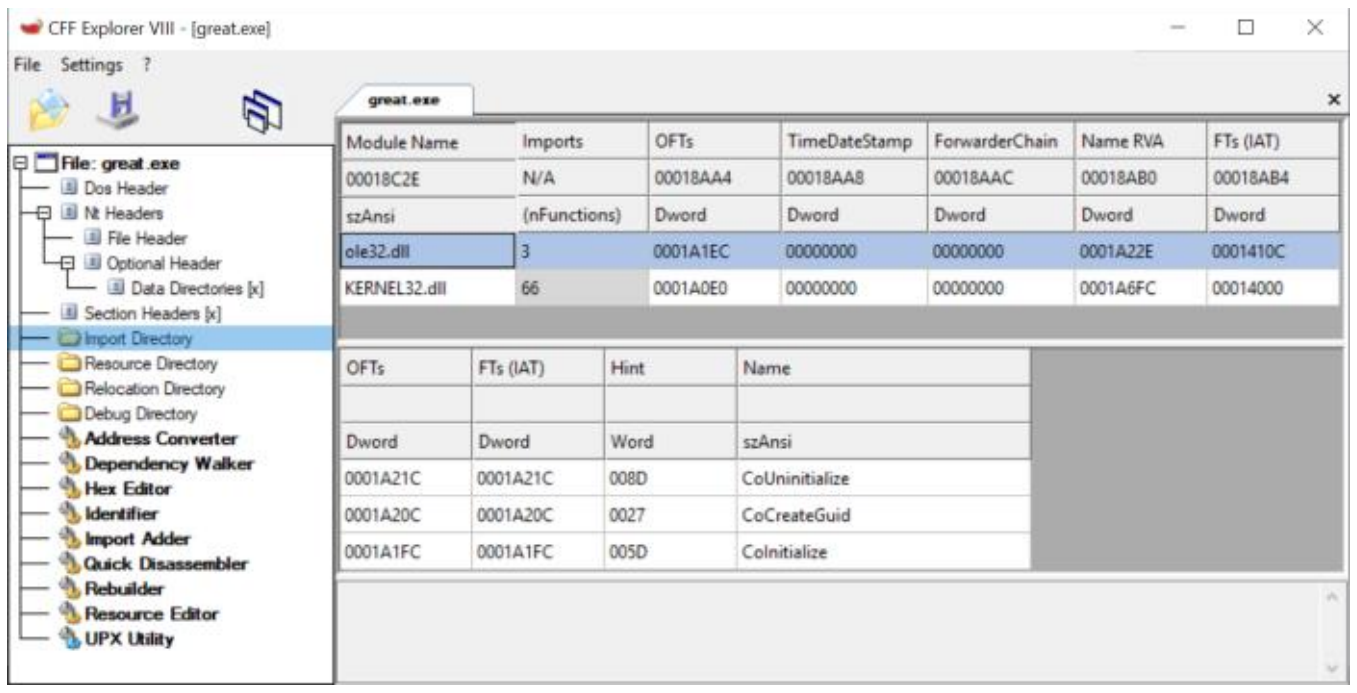

Figure 1: IMAGE_IMPORT_DESCRIPTOR for symbols imported from ole32.dll

If you run great.exe via WinDbg, notice that the loader has stored the address of ole32!CoInitialize in the thunk great!__imp_CoInitialize. Therefore, the instruction call dword ptr [great!__imp__CoInitialize] results in the execution of the function ole32!CoInitialize.

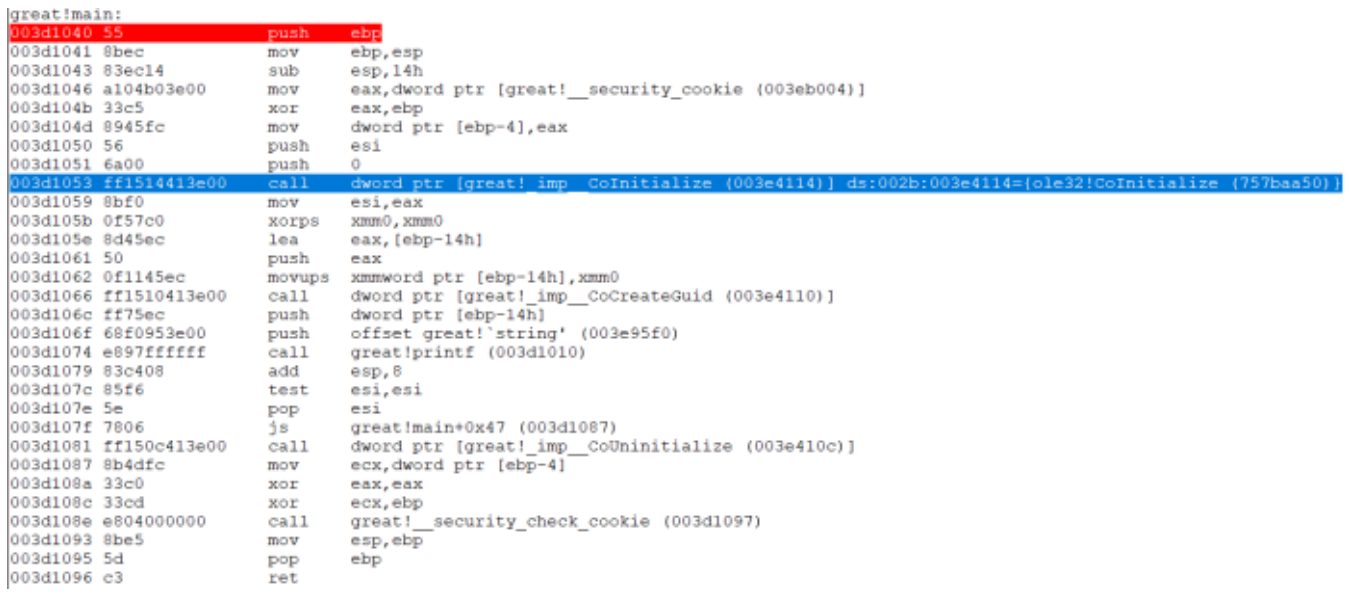

*Figure 2: __imp__CoInitialize thunk holds address of ole32!CoInitialize*

This is the disassembly of great.exe with symbols in IDAPro. Notice the call instructions shown as they correctly refer to the imports.

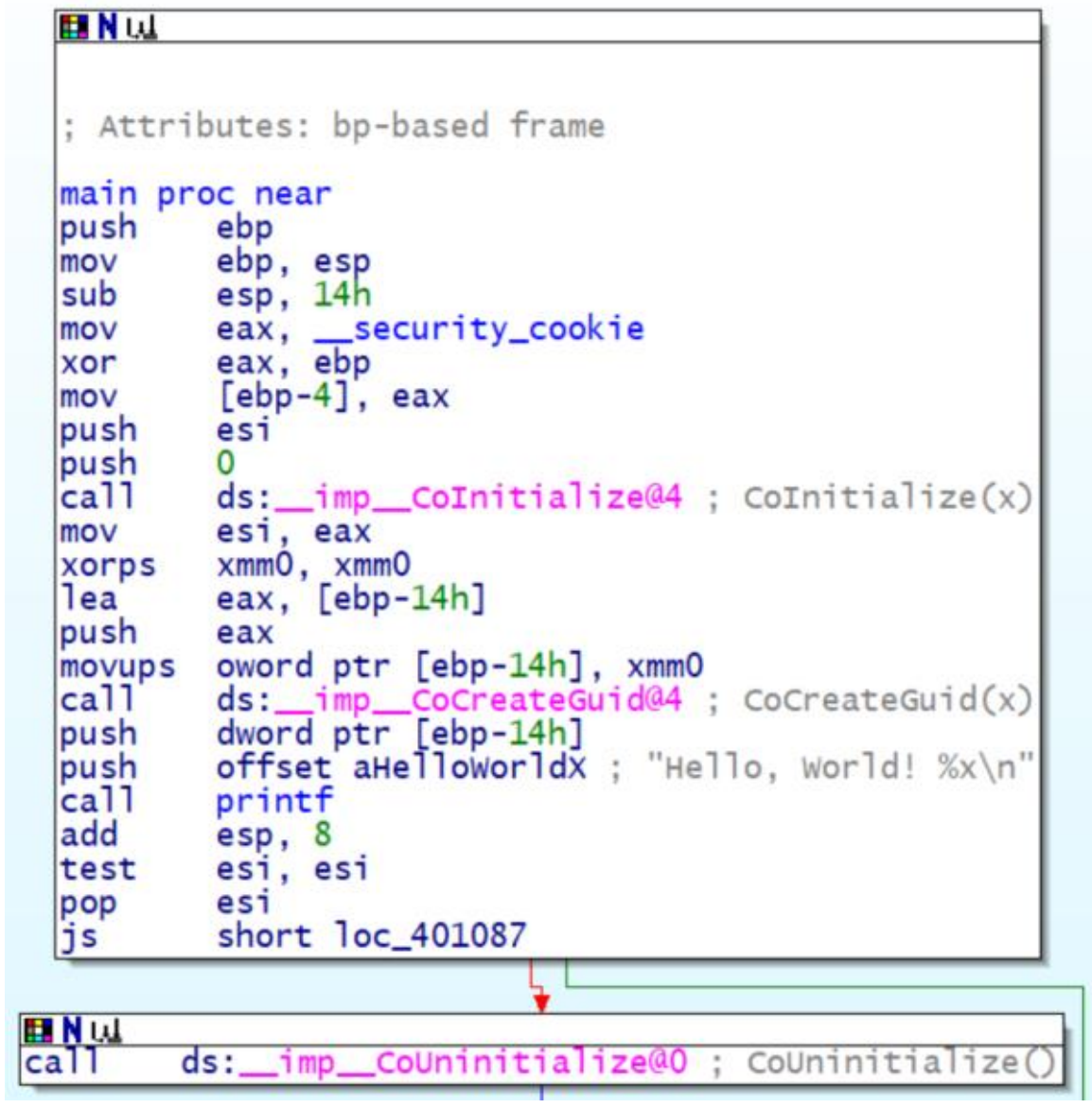

Figure 3: Disassembly of great.exe with program database in IDAPro

Now, using the Breaking-Imphash algorithm described, we transform great.exe, resulting in transformed.exe with imphash

c489844a22db2fe8ec5608c8de7473bd. In IDAPro, given that transformed.exe references great.exe's program database file for symbols, the view is interesting. Observe that CoCreateGuid resides where CoInitialize is expected, CoUninitialize resides where CoCreateGuid is expected, and CoInitialize resides where CoUninitialize is expected.

The IMAGE_IMPORT_DESCRIPTOR for ole32.dll in transformed.exe is shown in CFF Explorer in Figure 5. Notice that the order now is CoCreateGuid, CoInitialize, and CoUninitialize. The order of the symbols imported from kernel32.dll was also changed and this is in Appendix A.

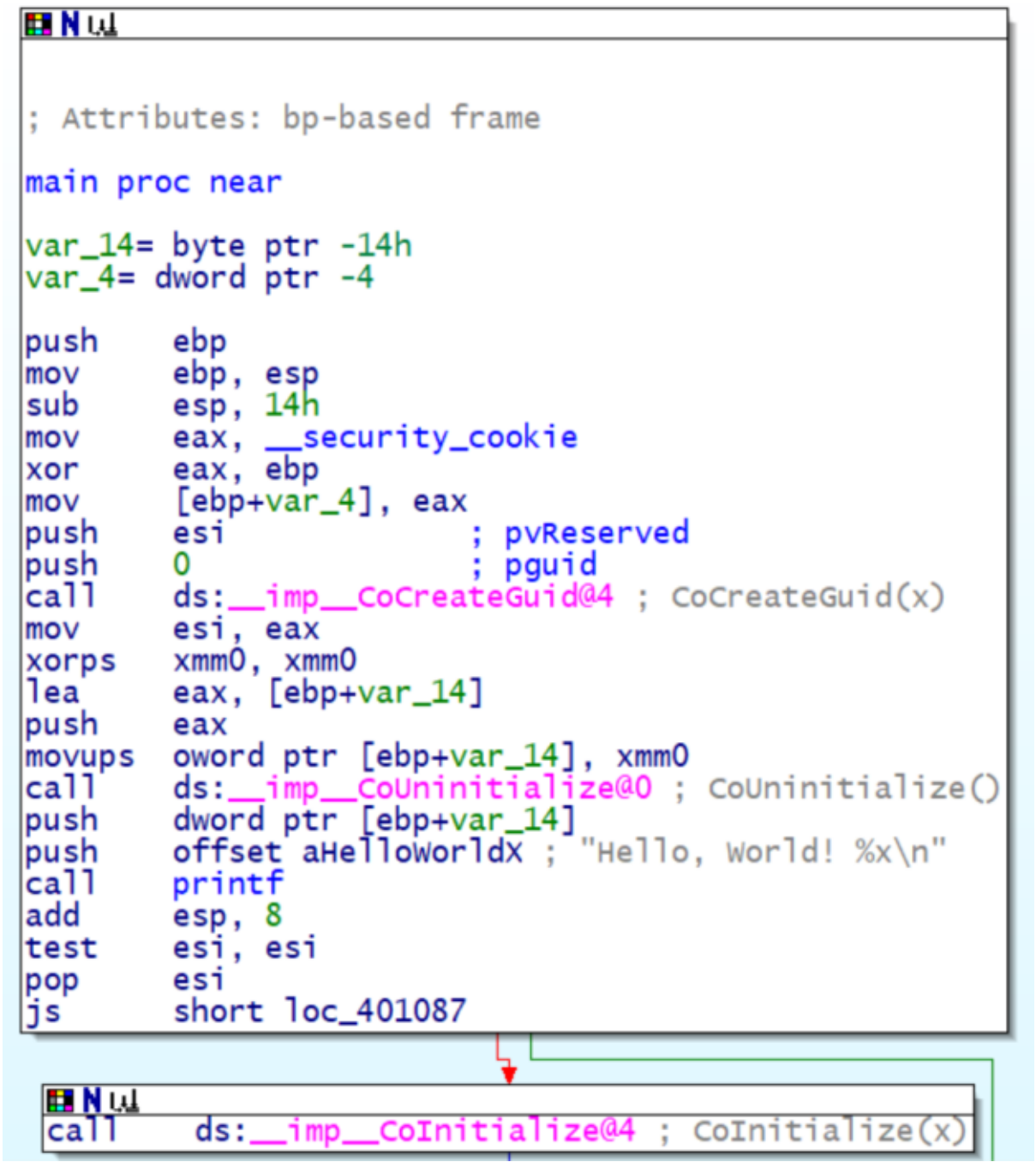

Figure 4: Disassembly of transformed.exe with great.exe's program database

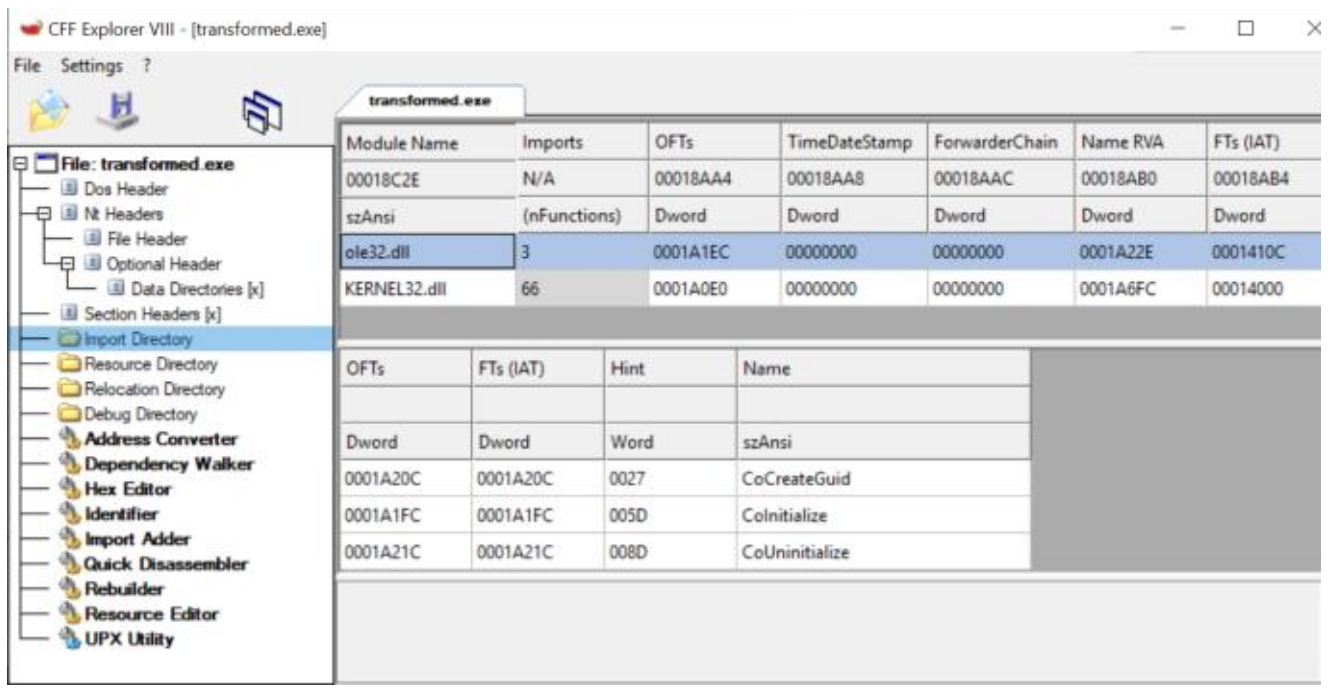

Figure 5: transformed.exe's IMAGE_IMPORT_DESCRIPTOR for symbols imported from ole32.dll

Now, if we look at the disassembly in WinDbg, notice that the loader is storing ole32!CoInitialize in the thunk which originally contained great!__imp_CoCreateGuid.

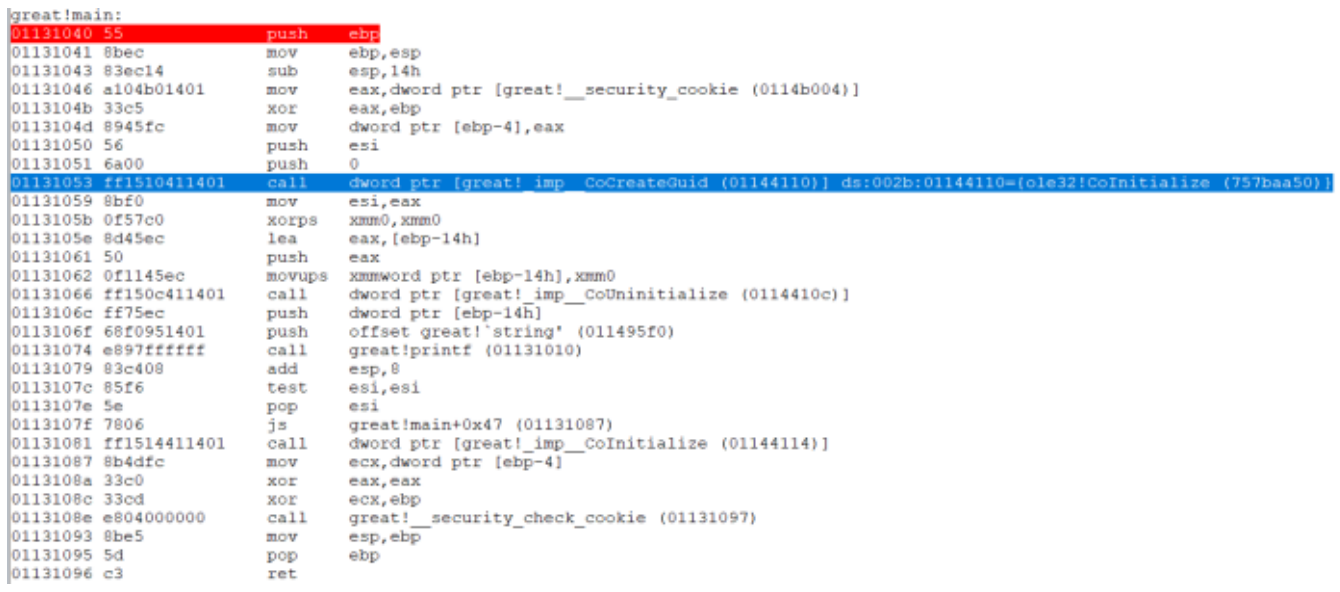

Figure 6: __imp__CoCreateGuid thunk holds ole32!CoInitialize

This great.exe example import 66 symbols from kernel32.dll and 3 from ole32.dll, and we can derive 69! different artifacts each with a unique imphash for this artifact. Note that 69! is larger than the 2128 (since imphash is MD5). That is, for great.exe, we can derive more artifacts than all available imphash values.

## IV. Conclusion

Breaking-Imphash algorithm is a feasible solution to generate derived PE files with imphash different from the original PE file. The algorithm requires the operating system to not load the PE file at its preferred base address in order to perform relocation. Relocation of PE is standard: Operating systems perform relocations for security or if another module is already loaded at the preferred base address. However, relocation can be disabled, for example, if IMAGE_FILE_RELOCS_STRIPPED flag is set in the PE file. This specific case can also be accommodated: Write an algorithm to inspect the .text section and identify all the call dword ptr instructions (ff 15 xx xx) that reference thunks. Replace these xx xx xx xx addresses with the updated thunk offsets. This approach is slightly more involved but not infeasible.